\documentclass[useAMS,usenatbib]{mn2e}
\usepackage{color}

\newcommand{\Lsun}{L$_{\odot}$}
\newcommand{\Msun}{M$_{\odot}$}

\newcommand{\avg}[1]{\left< #1 \right>}

\title[The effect of spiral arms on star formation]{The Effect of Spiral Arms on Star Formation in the Galaxy}
\author[T. J. T. Moore et al.]{T. J. T. Moore$^{1}$\thanks{E-mail:
tjtm@astro.livjm.ac.uk (TJTM)} J.S. Urquhart$^{2,3}$ L.K. Morgan$^1$ and M.A. Thompson$^4$\\
$^{1}$Astrophysics Research Institute, Liverpool John Moores University, Twelve Quays House, Egerton Wharf, Birkenhead CH41 1LD, UK\\
$^{2}$Australia Telescope National Facility, CSIRO Astronomy and Space Science, PO Box 76, Epping NSW 1710, Australia\\
$^{3}$Max-Planck-Institut f\"ur Radioastronomie, Auf dem H\"ugel 69, 53121 Bonn, Germany \\
$^{4}$Centre for Astrophysics Research, Science \& Technology Research Institute, University of Hertfordshire, College Lane, Hatfield,\\ Herts, AL10 9AB, UK 
}
\begin{document}

\date{Accepted 2012 July 17. Received 2012 July 14; in original form 2012 February 28}

\pagerange{\pageref{firstpage}--\pageref{lastpage}} \pubyear{2012}

\maketitle

\label{firstpage}

\begin{abstract}
We have examined the ratio between the integrated luminosity of massive young stellar objects detected by the Red MSX Source (RMS) survey and the mass of molecular clouds in the Galactic Ring Survey region, as a function of Galactocentric radius.  The results indicate that 60--80\% of the observed increases in the star-formation rate density associated with spiral-arm features are due to source crowding within the arms.  Of the remainder, most of the increase in the inner Sagittarius arm is due to an enhancement in the simple star-formation efficiency, i.e.\ in the number of RMS sources per unit molecular gas mass.  In the inner Perseus arm, the residual increase is due to a higher than average mean source luminosity, which implies a top-heavy IMF, and this is entirely due to the presence, in the GRS region, of the W49 star-forming complex, which appears to be exceptional in its nature.  The results also suggest that there is little or no increase in the star-formation efficiency on kiloparsec scales in the Scutum tangent region which includes W43.  We discuss the possible role played by the spiral arms in influencing the star-formation efficiency and conclude that the most likely mechanisms are related to orbit crowding within the arms.  
\end{abstract}


\section{Introduction}

The role that spiral arms play in affecting the process and efficiency of star formation in galaxies is not yet clear.  They may simply organise the gas of the interstellar medium (ISM), along with its molecular clouds, 
into regions of higher density, thus increasing the local star-formation rate density.  
Increases in star-formation efficiency (SFE) could result from such crowding, via a rise in the incidence or strength of local feedback, e.g. from earlier massive star formation, inducing additional star formation. 
On the other hand, spiral arms may be more direct triggers of star formation.  They may raise the probability of cloud-cloud collisions or increase the efficiency with which molecular clouds form from the neutral gas, via the shocks expected as the ISM gas enters the arm, or by altering the internal state, i.e.\ the average mass, velocity dispersion or lifetime of molecular clouds so as to affect their internal SFE.

\citet{HT98} found evidence, in H{\sc i} and CO data in W3/4/5, that the molecular fraction of the gas content in the outer Perseus spiral arm was around ten times higher than in the inter-arm regions on the line of sight.  This result implies that spiral arms dramatically raise the efficiency with which molecular clouds are produced out of the neutral gas.  In turn, this suggests that spiral density waves trigger additional star formation, via the creation of new molecular clouds, as modelled by, e.g., \citet{Dobbs06}.  In contrast, recent observations of two external spiral galaxies by \citet{Foyle} indicate that the H$_2$/H{\sc i} fraction and the infrared- and UV-traced SFE are not significantly enhanced in spiral arms relative to the inter-arm gas.  Also, \citet{leroy} conclude that the fraction of GMCs formed from H{\sc i} is governed by ISM physics acting on relatively small scales, i.e.\ the H$_2$ formation/destruction rate balance and stellar feedback.  \citet{krumholz09} predict that, except in starburst conditions, molecular-cloud properties are dominated by internal radiative feedback and not the environment.

\citet{Dobbs11} suggest that spiral arms are mainly organising features, 
whose main effect on the interstellar medium is to delay and crowd the gas, which is deflected from circular orbits while within the arm.  The SFR is increased indirectly by enabling longer-lived and more massive giant molecular clouds.  \citet{Roman10} conclude from observations that molecular clouds within spiral arms are more long-lived than in the inter-arm gas, which would lengthen the star-formation timescale for a typical cloud, increasing the SFE as a result. If larger or denser clouds are formed, it may be significant.  \citet{krumholz10} predict that the column density of clouds affects the mass function of clusters that form within them via radiative heating which suppresses fragmentation in the higher-column clouds but does not significantly affect the overall SFR/E.

It is also likely that spiral arms are different from each other \citep[e.g.][]{Benjamin05}.  
Also, the inner and outer portions of spiral arms may influence star formation differently.  The entry
shock experienced by the ISM gas entering a spiral arm should only exist inside the co-rotation radius, where there is a differential velocity between the spiral pattern speed and the orbital rotation speed of the
galactic ISM.  Outside this radius (thought to be just beyond the Solar circle in the Milky Way \citep{lepine}, supernovae may be the dominant mechanism determining the state of the ISM and, hence, star formation (e.g.\ \citealp{kobayashi}, \citealp{dib}).  The presence of the Galactic bar may also affect star formation, especially near the bar ends where the 
pattern rotates at the same speed as the adjacent orbiting gas and where crowding between orbits in the bar potential and external circular orbits may create a higher cloud collision rate.

Despite the abundance of clues, or perhaps because of it, our knowledge of the effect of spiral arms on the star-formation rate or efficiency is still poorly developed and the main questions about the relationship between large-scale Galactic structure and star formation remain unanswered.  

This paper reports the results of an investigation into the dependence of the efficiency of star formation on Galactocentric radius and proximity to Galactic spiral arms. The SFE is estimated by the ratio of the luminosity produced by infrared-selected, massive young stellar objects (YSOs) and H{\sc ii} regions to the CO-traced mass in molecular clouds.  In Sections 2 and 3 we describe the data and present the results. In Section 4 we discuss the findings and the implications for understanding the effect of Galactic structure on star formation.

\section[]{Data}

The data used for this study consist of the sample of molecular clouds extracted from the $^{13}$CO J=1--0 BU/FCRAO Galactic Ring Survey (GRS: \citealp{Jackson}) by \citet{Rathborne09} which have been 
positionally matched to mid-infrared-selected, massive young stellar objects 
and H{\sc ii} regions from the Red MSX Source (RMS) survey by \citet{Urquhart11}.  
The result is a sample of molecular clouds with IR-detected high-mass star 
formation, as well as a complementary sample of clouds without RMS detections, all with 
kinematic distances, estimated cloud masses and total source luminosities. Distance 
ambiguities were resolved mainly by \citet{Roman09} with additional determinations by 
\citet{Urquhart11}, in which details of the construction of the matched sample can also be found.  
The GRS cloud-mass estimates were revised upwards by a factor of several in a re-analysis by \citet{Roman10}. The spatial coverage of the data is $17^\circ\!.9 < l < 55^\circ\!.7$ and $|\,b\,| \le 1^\circ$, and the velocity range is $-5$ to 135\,km\,s$^{-1}$ for $l < 40^{\circ}$ and $-5$ to 85\,km\,s$^{-1}$ at $l > 40^{\circ}$, both set by the coverage of the GRS.  The RMS sample is complete to $L_{\rm bol} > 10^4$\,L$_\odot$ out to a heliocentric distance of $\sim$14\,kpc and covers a Galactocentric radius ($R_{\rm GC}$) range of 2.5 to 8.5\,kpc. The combined sample consists of 176 RMS sources with 123 GRS cloud associations and 423 GRS clouds with no matching RMS detection above $10^4$\,L$_\odot$.

Only GRS clouds with masses above $5\times10^4$\,\Msun\ are included in the sample.  This is a more conservative limit than adopted by \citet{Roman10} ($1.1\times10^4$\,\Msun).  The sample was also limited to sources with heliocentric distances greater than 2\,kpc, in order to remove local sources that might affect the results at $R_{\rm GC} \simeq 8$\,kpc.

Source luminosities were obtained by constructing SEDs from various public data sources
and fitting them with the YSO model fitter of \citet{robitaille}.  Luminosities are effectively bolometric
although necessarily dominated by infrared data (see \citealp{Mottram11b} for details).

\section{Analysis and Results}

\begin{figure}
 \vspace{6.8cm}
\includegraphics{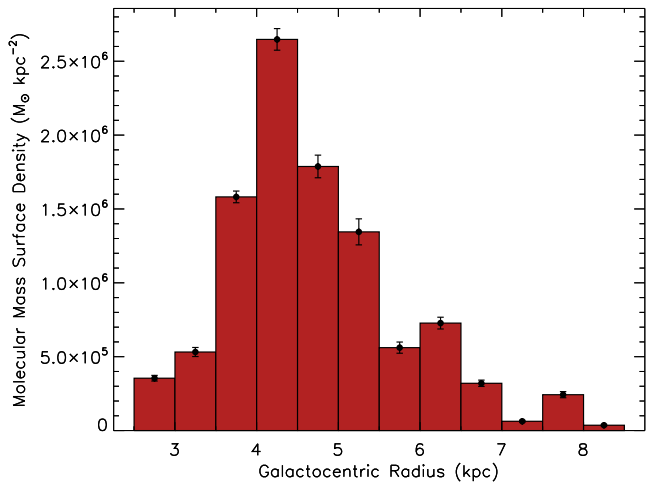}
 \caption{
Galactic surface density of molecular mass in the GRS region as traced by clouds in the catalogue of \citet{Rathborne09}, using the revised masses of \citet{Roman10} and a lower limit of $5 \times 10^4$\,\Msun per cloud.  The $x$ scale is Galactocentric radius and the bin size is 0.5\,kpc. Errors are derived from the mass uncertainties listed by \citet{Roman10}.
} 
\label{COmass}
\end{figure}

\begin{figure}
 \vspace{6.8cm}
\includegraphics{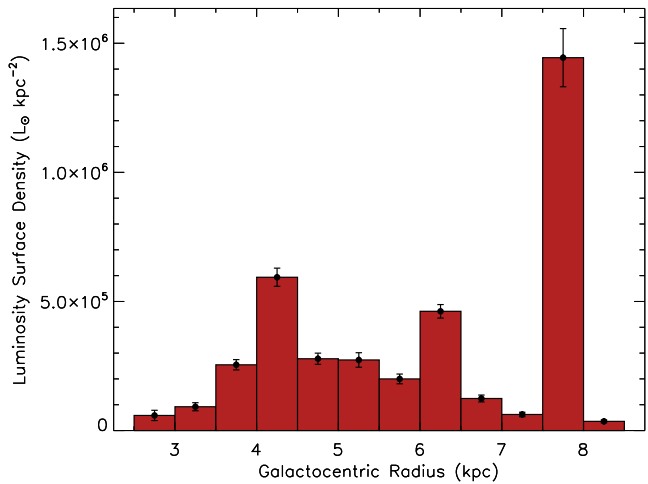}
 \caption{
The Galactic surface density of  RMS source luminosity above the completeness limit of $10^4$\,\Lsun, as a function of Galactocentric radius.  Sources within 2\,kpc of the Sun and those associated with GRS clouds of $M<5 \times 10^4$ \Msun\ have been removed. The error bars are based on the average uncertainty on the total IR luminosity of 34\% estimated by \citet{Mottram11}
} 
\label{Lperarea}
\end{figure}

\begin{figure}
\vspace{6.8cm}
\includegraphics{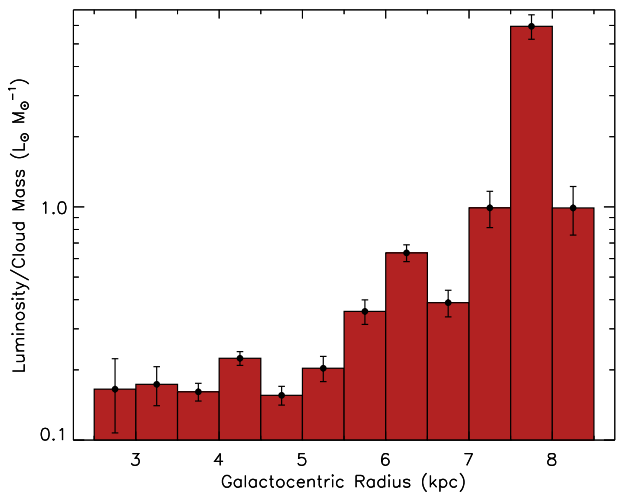}
 \caption{
The ratio of integrated RMS source luminosity to mass in GRS clouds ($L_{\rm bol}/M_{\rm CO}$) as a function of Galactocentric radius. The $y$ scale is plotted logarithmically in order to show detail at low levels.
 } 
\label{SFEbasic}
\end{figure}

Figure \ref{COmass} shows the mean mass surface density of molecular gas in $^{13}$CO-traced GRS clouds ($\Sigma_{M_{\rm CO}}$), selected as described in Section 2 above, as a function of $R_{\rm GC}$.  This result is the same as that in Fig.\ 8 of \citet{Roman10} and has been seen in a number of previous works (e.g.\ \citealp{lbx84}; \citealp{liszt93}).  There is a large peak in $\Sigma_{M_{\rm CO}}$ at $R_{\rm GC} = 4-5$\,kpc.  This region corresponds to the Scutum spiral-arm origin and inner tangent and the end of the Galactic bar at Galactic longitude $l \sim 30^{\rm o}$, where the massive star-forming regions W43 and G29.96 are located (see, e.g., Bally et al., 2010, \citealp{Luong11}).  The Scutum arm material occupies a relatively large range of velocities and kinematic distances (\citealp{Luong11}, \citealp{eden}) but a much narrower range in $R_{\rm GC}$ and so is a well defined feature using this scale.

At radii less than $\sim$4\,kpc, $\Sigma_{M_{\rm CO}}$ falls rapidly to comparatively very low values.  At larger radii, $\Sigma_{M_{\rm CO}}$ declines slightly less steeply, with two additional peaks at $R_{\rm GC}$ = 6--6.5 kpc and at 7.5--8 kpc.  These two zones correspond to the average radii of the inner segments of the Sagittarius and Perseus spiral arms, respectively, as they pass through the GRS region.

Figure \ref{Lperarea} shows the integrated luminosity of RMS MYSOs and H{\sc ii} regions per unit Galactic surface area ($\Sigma_{L_{\rm bol}}$) as a function of $R_{\rm GC}$.   
Three peaks can again be identified in the distribution at 4--4.5 kpc, 6--6.5 kpc and 7.5--8 kpc, corresponding to those seen in the $\Sigma_{M_{\rm CO}}$ 
distribution of Figure \ref{COmass}.   This time, however, the contrast between the three features is less marked and the contrast with the background levels is greater.  The background level of $\Sigma_{L_{\rm bol}}$ also falls steadily with radius beyond $R_{\rm GC} \simeq 5$\,kpc, but less steeply than the mass surface density. These features were also seen, in the same data, by \citet{Urquhart11}.

\begin{figure}
\vspace{14cm}
\includegraphics{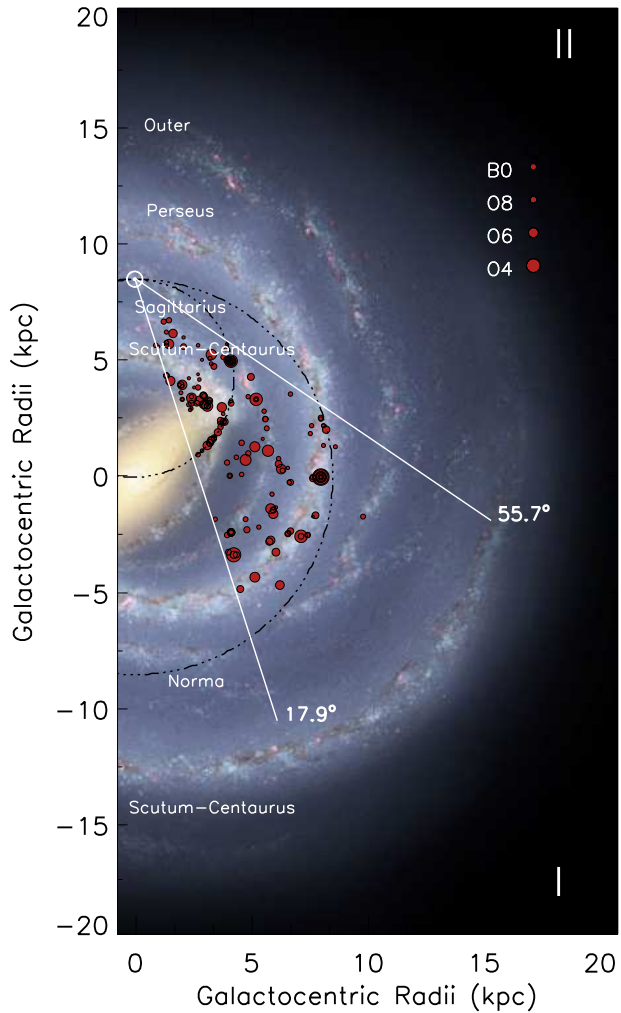}
 \caption{
The distribution of RMS YSOs that have assigned distances within the GRS region, superimposed on part of the sketch plan of the Galaxy by Robert Hurt of the {\em Spitzer} Science Center, in consultation with Robert Benjamin.  The key demonstrates the approximate luminosity of each RMS source.  The figure is adapted from \citet{Urquhart11}
 } 
\label{galplan}
\end{figure}

Figure \ref{SFEbasic} presents the ratio of the total luminosity $L_{\rm bol}$ associated with RMS sources to total molecular mass $M_{\rm CO}$ in GRS clouds in each radial bin.  From $R_{\rm GC} = 2.5$ to 5.5\,kpc, $L_{\rm bol}/M_{\rm CO}$ has a low value ($\sim 0.18$\,\Lsun/\Msun) and is nearly flat and featureless.  At $R_{\rm GC} > 5.5$\,kpc, $L_{\rm bol}/M_{\rm CO}$ begins to rise and goes through two discrete, significant peaks, the first at $R_{\rm GC}$ = 6--6.5\,kpc, where the value rises to $\sim 0.63$\,\Lsun/\Msun, 70\% higher than the adjacent bins, and a much stronger one of $\sim 5.9$ \Lsun/\Msun\ at $R_{\rm GC}$ = 7.5--8\,kpc, around six times larger than in the bins either side.  The latter two features again correspond to the radii of the Sgr and Per arms, respectively (see Figure \ref{galplan}).

The GRS survey does not cover the two spiral arms beyond Perseus which are the Norma arm and the distant, outer Scutum-Centaurus arm recently detected by Dame \& Thaddeus (2011).  Although at least partly within its spatial range, these are both outside the Solar circle and so at negative relative velocities, and thus not detected by the GRS.  The outer Scutum-Centaurus arm also lies partly above the latitude range of the GRS, following the upward warp of the Plane.

\section{Discussion}

\subsection{Star-formation efficiency vs the massive YSO luminosity function}

The quantities plotted in Figures \ref{COmass} and \ref{Lperarea}, $\Sigma_{M_{\rm CO}}$ and $\Sigma_{L_{\rm bol}}$, depend on the number density of molecular clouds as well as on any physical differences in the sources.  Their ratio, $L_{\rm bol}/M_{\rm CO}$ (Figure \ref{SFEbasic}), is independent of source crowding and therefore reveals differences in at least the outcome of the star-formation process as a function of Galactic radius.  $L_{\rm bol}/M_{\rm CO}$ is determined by a combination of the mean star-formation efficiency (SFE), which is the star-formation rate (SFR) per unit gas mass, integrated over the relevant timescale, and the luminosity function (LF) of the massive young stars.  If the SFE is high enough to cause significant depletion of the molecular gas mass reservoir in a star-formation time, its dependence on the SFR may become non-linear.

An increase in the value of $L_{\rm bol}/M_{\rm CO}$ can therefore be produced by one or more of the following: a rise in the SFR per unit gas mass, a shallower LF (i.e.\ weighted towards high-luminosity sources), or a long time period.  The timescale sampled by the data is limited to the lifetimes of those evolutionary stages traced by the RMS survey.  These are the massive YSO (MYSO) and compact H{\sc ii}-region stages, the durations of which have both been determined to be $<5\times 10^{5}$\,yr by \citet{Mottram11}.  This timescale is short enough, compared to the lifetime of a molecular cloud, that the data provide a snapshot of the current star formation.  We therefore consider that it is not necessary to account for differing lengths of time over which star formation might have continued.  

Where the LF of forming stars is invariant, the ratio $L_{\rm bol}/M_{\rm CO}$ is simply a measure of the current SFE.  In this case, Figure \ref{SFEbasic} indicates a significantly higher SFE in the two outer spiral arms (Sgr and Per) compared to that in the region with $R_{\rm GC} < 6$\,kpc and, importantly, relative to that in the neighbouring inter-arm clouds.  This suggests that physical conditions in the molecular clouds within these arms are altered in some way.  Additionally, Fig.\ \ref{SFEbasic} shows that the average SFE is significantly higher in the inner Perseus arm compared to the inner Sgr arm, and so that conditions for star formation may be different from arm to arm.  

Despite very strong peaks in both 
$\Sigma_{M_{\rm CO}}$ and $\Sigma_{L_{\rm bol}}$  at $R_{\rm GC}$ = 4--5\,kpc (Figures \ref{COmass} and \ref{Lperarea}), $L_{\rm bol}/M_{\rm CO}$ in this inner region is low and nearly constant.  
These inner radii contain the Scutum spiral-arm tangent and the near end of the Galactic bar, including the massive star-forming regions W43 and G29.96. The former has been described as starburst-like (e.g.\ \citealt{bally}, \citealt{Luong11}), but Figure \ref{SFEbasic} implies that the intense concentration of star formation found there is largely the result of the huge amount of molecular gas along that line of sight and not to a significantly elevated SFE.   This result is consistent with those of \citet{eden}, who find no significant difference between the mass fraction of dense clumps in the molecular clouds associated with W43 and that in the foreground and background clouds on the same line of sight.   

\begin{figure}
 \vspace{6.8cm}
\includegraphics{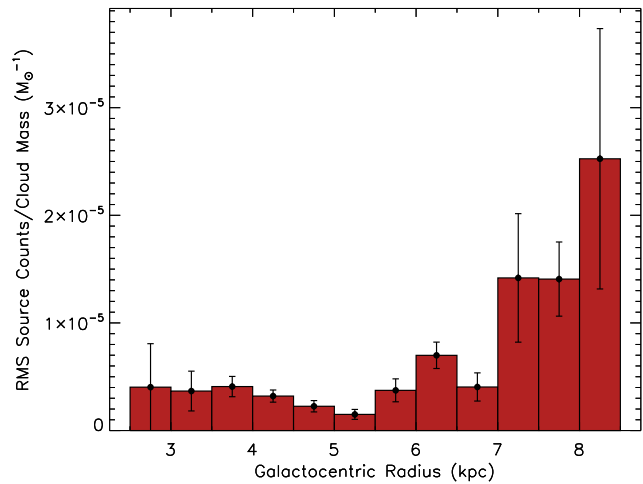}
 \caption{
The number of RMS MYSO sources above $10^4$\,\Lsun\ per unit molecular gas mass traced by GRS clouds ($N_{\rm RMS}/M_{\rm CO}$).  The error bars represent Poisson errors on the RMS source counts combined with uncertainties on mass estimates from \citet{Roman10}, with the latter adjusted for a distance uncertainty of 1\,kpc.
} 
\label{NperCOmass}
\end{figure}

While there is little strong observational evidence to support the hypothesis that the IMF of massive stars is sensitive to the initial conditions for star formation \citep{Bastian10}, it is still possible that $L_{\rm bol}/M_{\rm CO}$ may also depend on the LF of the massive young stars that are forming.  In such a case, the same SFE but with a flatter IMF would yield a higher value of $L_{\rm bol}/M_{\rm CO}$.  Since the latter is made up of the number of YSOs formed per unit cloud mass and the LF of those YSOs, we can partly separate the two effects by examining independently the numbers of RMS sources per unit cloud mass, $N_{\rm RMS}/M_{\rm CO}$, and the RMS source LF.  

Figure \ref{NperCOmass} shows $N_{\rm RMS}/M_{\rm CO}$ as a function of $R_{\rm GC}$.  
Against a rising background value beyond $R_{\rm GC} \simeq 5$\,kpc, there is a small peak in $N_{\rm RMS}/M_{\rm CO}$ at 6.0--6.5 kpc, the radius of the Sgr arm.  The apparent significance of this peak is barely 3 sigma because of the large Poisson error bars, but it represents an increase of around 70\% over the neighbouring points and is enough to account for the peak in $L_{\rm bol}/M_{\rm CO}$ at the same radius in Figure \ref{SFEbasic}.   There is therefore no need to suspect any change in LF associated with the Sgr arm and an increase in the number of YSOs produced per unit gas mass, i.e., a simple increase in the SFE, appears to be a sufficient explanation. 
In contrast, Figure \ref{NperCOmass} shows no evidence of any rise in $N_{\rm RMS}/M_{\rm CO}$ in the $R_{\rm GC} = 7.5-8.0$-kpc bin, and so a simple increase in SFE does not explain the large peak in $L_{\rm bol}/M_{\rm CO}$ in the Per arm seen in Fig.\ \ref{SFEbasic} and we need to look for changes in the LF as a function of $R_{\rm GC}$.

\begin{figure}
 \vspace{6.8cm}
\includegraphics{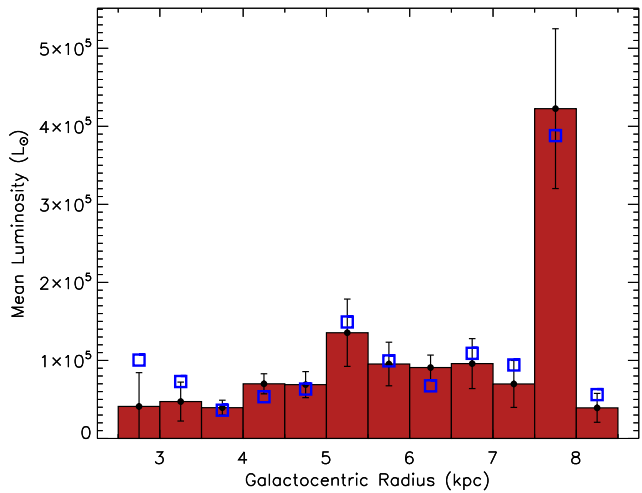}
 \caption{
The mean luminosity of RMS sources ($\avg{L_{\rm bol}}$) with $L_{\rm bol} \ge 10^4$\,\Lsun, as a function of Galactic radius, i.e.\ the total luminosity divided by the number of sources per bin.  The blue squares show the effect of correcting for a $N^{1/3}$ dependence in $\avg{L_{\rm bol}}$ (see text).
} 
\label{meanluminosityfig}
\end{figure}

The LF of MYSOs has been determined for the whole Galaxy by \citet{Mottram11}, but the present sample is not large enough to generate an explicit LF for each 0.5-kpc $R_{\rm GC}$ bin.  However, we can crudely examine the LF as a function of $R_{\rm GC}$ via the mean luminosity $\avg{L_{\rm bol}}$, both explicitly and  and by statistical tests for differences in the luminosity distributions.

Figure \ref{meanluminosityfig} shows the value of $\avg{L_{\rm bol}}$, i.e.\ the total $L_{\rm bol}$ divided by the number of RMS sources in each $R_{\rm GC}$ bin.  This time we see a significant increase by a factor of $\sim$7 at the 8-kpc radius of the Per arm but no change at the Sgr arm (6.5\,kpc) and a flat or slowly rising value ($\sim (4-7)\times10^4$\,\Lsun) between 3 and 5 kpc. Some caution is required here, since the distribution of luminosities $N(L_{\rm bol})$ is a power law with a lower cutoff imposed by incompleteness and an upper cutoff at $N=1$, due to integer values of $N$.  Hence $L_{\rm bol}/N$ is an increasing function of $N$, i.e.\ the sample size.  If the power-law exponent is $-1.5$, then $\avg{L_{\rm bol}} \propto N^{1/3}$ exactly.  The normalised correction for this bias is shown in Figure \ref{meanluminosityfig} but is obviously smaller than the $\sqrt{N}$ uncertainties on $N$.  The true LF of RMS-traced massive star-forming regions may be shallower this (\citealp{Mottram11}) and the severity of the bias rises sharply for exponents more positive than $-1$. However, since there are fewer RMS sources in the 7.5-8.0 kpc Perseus-arm bin (19) than in the Sagittarius- and Scutum-arm bins (36 and 33, respectively), such a bias cannot explain the much higher $\avg{L_{\rm bol}}$ in the former. This result therefore implies a significant flattening of the MYSO luminosity function in the Per arm.

Two statistical tests of the distribution of $L_{\rm bol}$ were performed, looking for differences between the RMS sources in the Perseus-arm subsample ($R_{\rm GC} = 7.5 - 8.0$\,kpc) and the rest of the sample.  The Mann-Whitney U test, which is mainly sensitive to displacements (i.e.\ differences in the means) between two samples which are not normally distributed, produced a probability that the two samples come from the same distribution of $p=0.0004$.  The Kolmogorov-Smirnov test for general differences in the two distributions resulted in $p = 0.028$.  The null hypothesis (that the samples come from the same distribution) can therefore be rejected at least at the 2-$\sigma$ level.

\begin{figure}
 \vspace{6.8cm}
\includegraphics{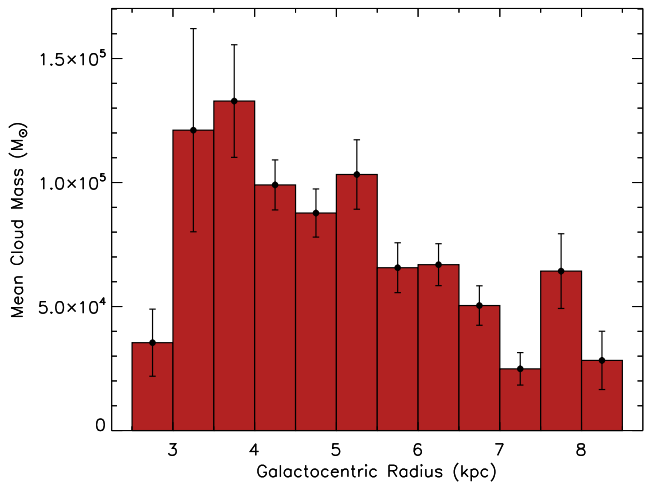}
 \caption{
The average GRS cloud mass as a function of 
Galactocentric distance.  
} 
\label{avemassfig}
\end{figure}

As well as the discrete features associated with the spiral arms, Figures \ref{SFEbasic} and \ref{NperCOmass} show a gradual increase in the baseline value of both $L_{\rm bol}/M_{\rm CO}$ and $N_{\rm RMS}/M_{\rm CO}$ by factors of about 5 between $R_{\rm GC} = 4$ and 8\,kpc.  This gradient can be largely explained by Figure \ref{avemassfig}, which  shows a steady decrease, beyond $R_{\rm GC}$ = 4\,kpc, in the mean molecular-cloud mass $\avg{M_{\rm CO}}$ by a similar factor.   The latter result was also seen by \citet{Roman10} and will be discussed in more detail in a forthcoming paper.  
Figure \ref{avemassfig} also shows a discrete peak in $\avg{M_{\rm CO}}$ at the radius of the Per arm, but none at the Sgr arm or the Scutum tangent region.  

\subsection{The effect of individual sources within the arms}

The peaks in $L_{\rm bol}/M_{\rm CO}$ at $R_{\rm GC} \sim 6$ and 8\,kpc are both due to the presence of individual star-forming complexes with very high local values of $L_{\rm bol}/M_{\rm CO}$ within the corresponding spiral arms.  At $\sim$6 kpc, W51A and W51B are associated with the GRS cloud G049.49--00.41 which has a Galactocentric distance of 6.5\,kpc (\citealp{Rathborne09}).  The mass of this cloud is determined to be $(1.8 \pm 0.5)\times10^5$\,\Msun\ (\citealp{Roman10}) and its integrated luminosity in RMS sources with $L>10^4$\,\Lsun\ is estimated to be $1.37\times10^6\,$\Lsun, giving $L_{\rm bol}/M_{\rm CO} = 7.6 \pm 2.3$\,\Lsun\,M$_{\odot}^{-1}$.  Nine RMS sources are associated with this cloud \citep{Urquhart11}, seven of which are above the $10^4$-\Lsun\ completeness limit.  Just three of these sources have luminosities above $10^5$\,\Lsun\ (RMS\,49.4903--00.3694, 49.5373--00.3929 and 49.4564--00.3559).

\citet{kang} estimate the total molecular gas mass in W\,51 as $2.3\times 10^5$\,\Msun\ in at least 8 clouds while \citet{carpenter} obtain $1.2\times 10^6$\,\Msun.  \citet{harvey} set the total luminosity at $3\times10^6$\,\Lsun, using an assumed distance of 7\,kpc.  The distance has been estimated by trigonomentric parallax at $5.4\pm 0.3$ kpc \citep{sato}, which implies $R_{\rm GC} = 6.3$\,kpc.  At this distance, the Harvey et al.\ luminosity estimate becomes $1.8\times 10^6$\,\Lsun.  The molecular gas content of W51 has been studied in detail by \citet{parsons}.

At $R_{\rm GC} \simeq 8$\,kpc, the well-known star-forming region W49A is associated with GRS cloud G43.19--00.01, whose mass is $(2.2\pm 0.3)\times10^5$\,\Msun\ and total RMS source luminosity is $6.87\times10^6$\,\Lsun. This gives $L_{\rm bol}/M_{\rm CO} = 32\pm 6$\, \Lsun\,M$_{\odot}^{-1}$.  Nine RMS MYSOs are associated with this cloud, all with luminosity above $10^4$\,\Lsun. Five of these have $L>10^5$\,\Lsun\ and the integrated luminosity is dominated by two sources (RMS\,43.1679--00.0095 and 43.1650--00.0285) with $L_{\rm bol}>10^6$\,\Lsun.

\citet{roberts} suggest that W49A is a good Galactic analogue for an extragalactic starburst system, having
gas temperatures of 50--100\,K and densities $\sim$$10^6$\,cm$^{-3}$.  Its distance was determined from maser proper motions to be $11.4\pm1.2$\,kpc \citep{gwinn}. Its total luminosity was estimated at $6\times 10^6$\,\Lsun\ by \citet{harvey77} and at $>10^7$\,\Lsun\ by \citet{sievers} and \citet{dwt} and it contains more than a dozen compact H{\sc ii} regions.  It has been suggested that the 
high star-formation rate in W49 is the result of a cloud-cloud collision resulting from orbit crowding in the spiral arm \citep{serabyn}.  The total associated gas mass in several clouds in the region has been variously assayed at $1.7\times 10^6$\,\Msun\ \citep{miyawaki} and $1.1\times 10^6$\,\Msun\ \citep{williams}.

If we remove the GRS cloud containing the W49A RMS sources, the $R_{\rm GC}$ = 7.5--8.0-kpc bin then has integrated $L_{\rm bol}/M_{\rm CO} = 1.03\pm 0.15$\,\Lsun\,\Msun$^{-1}$, $N_{\rm RMS}/M_{\rm CO} = (0.9 \pm 0.3) \times 10^{-5}$\,\Msun$^{-1}$ and $\avg{L_{\rm bol}} = (1.1 \pm 0.4) \times 10^5$\,\Lsun, similar to the levels in the adjacent bins in each case (Figs \ref{SFEbasic}, \ref{NperCOmass} and \ref{meanluminosityfig}).  Removing the cloud associated with W51A \& B, the 6.0--6.5-kpc bin has $L_{\rm bol}/M_{\rm CO} = 0.38\pm 0.03$\,\Lsun\,\Msun$^{-1}$, $N_{\rm RMS}/M_{\rm CO} = (0.5 \pm 0.1) \times 10^{-5}$\,\Msun$^{-1}$ and $\avg{L_{\rm bol}} = (6.5\pm1.3) \times 10^4$\,\Lsun, which results in a slightly reduced peak in $L_{\rm bol}/M_{\rm CO}$, relative to the neighbouring bins.
This suggests that, without W51, the Sgr arm still has somewhat increased SFE while the Per arm has a flatter than normal LF solely due to the presence of the W49A massive YSOs.

\subsection{Implications for the effect of spiral-arm structure on star formation}

What does the foregoing analysis tell us about spiral arms as potential large-scale triggers of star formation? Figure \ref{Lperarea} shows increases in $\Sigma_{L_{\rm bol}}$ by factors of about 2.2, 2.9 and 30 in the Scu, Sgr and Per arm segments, respectively, relative to the adjacent inter-arm regions. Around 70\%, 60\% and 80\%, again respectively, of these rises are due to simple source crowding.  The remaining 20--40\% of the increase in $\Sigma_{L_{\rm bol}}$ is due to changes in the luminosity per unit cloud mass and may be associated with a physical effect on the clouds caused by the presence of the arms.

There is some indication in Figure \ref{avemassfig} that cloud masses may be larger in the Per spiral arm, while no similar increase is seen in the Sgr arm.  This may be an indication that clouds within the Per arm, or at least in W49, may be in an altered state.  If clouds are more massive, they may be less well supported on large scales.  Interestingly, the simulations of \citet{krumholz11} predict that global collapse of massive molecular clouds with radiative feedback should produce a top-heavy stellar IMF because of the suppression of fragmentation of the cloud, while accretion onto the forming protostars continues.  The overall SFE is not significantly affected.

Models by \citet{Dobbs08} predict that smaller clouds aggregate into larger ones within spiral arms and, as they leave the arms, clouds are subject to shear and lose molecular gas.  Dobbs, Burkert and Pringle (2011) suggest that, while more massive GMCs accumulate in the arms, there is no direct effect on the SFR, only an indirect influence via longer-lived and more strongly bound clouds. Pathological SF regions like W49A may be due to cloud-cloud collisions made more likely by longer cloud lifetimes and orbit crowding within spiral arms.

There is no clear evidence in the current results that can distinguish between changes in the clouds, caused by being within an arm, and increased triggered star formation caused by higher feedback between crowded star-forming regions.  However, the lack of significant increases in $L_{\rm bol}/M_{\rm CO}$ in the crowded Scutum tangent region suggests that feedback between clouds is not the dominant factor. 
\citet{dib12} found no significant connection between the shear resulting from the Galactic rotation and the star-formation efficiency in the GRS molecular clouds.

The Scutum tangent lies at the bar end where there should be co-rotation between the pattern speed of the bar and the ISM.  We might expect an altered star-forming environment in this region, at least from either the permanent presence of the bar potential and/or collisions between clouds in the circular orbits just outside the bar and gas following the $x_1$ orbits within it, as they reach the bar end.  On the other hand, since the CO mass surface density is seen to drop rapidly inside 3\,kpc, there may not be enough molecular gas in the bar for this to be a significant effect.

\section{conclusions}

Around 70\% of the increase in the SFR density in spiral arms is due to simple source crowding within
the arms.  The remaining $\sim 30$\% is the result of an increase in the luminosity coming from embedded massive YSOs, relative to the mass of molecular gas present. In the segment of the Sagittarius spiral arm included in the GRS data, this increase in $L_{\rm bol}/M_{\rm CO}$ is accounted for by a rise in the number of MYSOs per unit molecular gas mass, with no detected change in their mean luminosity compared to the nearby inter-arm areas.  This implies an increase in the basic SFE in molecular clouds in the Sgr arm, and this is probably caused by a higher SFR per unit gas mass, given the short timescales sampled by the RMS data, without any change in luminosity function.  In the Perseus arm segment, no increase in the number of RMS sources per unit cloud mass is detected, relative to the nearby inter-arm regions.  Instead, the rise in $L_{\rm bol}/M_{\rm CO}$ is wholly accounted for by an increase in the average luminosity of the MYSOs which implies a significant change in the average luminosity function, i.e.\ in the IMF.  Further, the changes in the Per arm are attributed wholly to the W49A star-forming complex.  If this were removed from the data, star formation in the Per arm would be similar to that in the inter-arm gas.  W49A appears to contain unusual star formation and may be genuinely starburst-like, while the major star-forming regions in Sgr (W51) and Scu (W43) may be part of a normal distribution of star-formation properties.

Compared to \citet{Foyle},who measured increases in SFR/$M$ of less than 10\% in the arms of external spiral galaxies, compared to inter-arm regions, our results show larger increases, of $\sim$30\% in the Sgr and Per arms.  In the Scutum tangent region, however, the enhancement in star-formation rate density is almost entirely due to source crowding.  This indicates variations within and between arms and that is is important to consider the scale to which such results correspond.

\section*{Acknowledgments}

This publication makes use of molecular line data from the Boston University-FCRAO Galactic Ring Survey (GRS). The GRS is a joint project of Boston University and Five College Radio Astronomy Observatory, funded by the National Science Foundation under grants AST-9800334, AST-0098562, \& AST-0100793.
This paper made use of information from the Red MSX Source survey database at $\mbox{www.ast.leeds.ac.uk/RMS}$ which was constructed with support from the Science and Technology Facilities Council (STFC) of the UK.  TJTM and LKM were supported, in part and whole, respectively, by STFC grant ST/001847/1.  The authors thank the referee, James Binney, for suggestions that have significantly improved the paper.

\label{lastpage}

\end{document}